# Screening and interlayer coupling in multilayer graphene field-effect transistors

Yang Sui* and Joerg Appenzeller*

School of Electrical and Computer Engineering and Birck Nanotechnology Center

Purdue University, West Lafayette, Indiana, USA 47907

sui@purdue.edu, appenzeller@purdue.edu

ABSTRACT: With the motivation of improving the performance and reliability of aggressively scaled nano-patterned graphene field-effect transistors, we present *the first* systematic experimental study on charge and current distribution in multilayer graphene field-effect transistors. We find a very particular thickness dependence for $I_{on}$, $I_{off}$, and the $I_{on}/I_{off}$ ratio, and propose a resistor network model including screening and interlayer coupling to explain the experimental findings. In particular, our model does not invoke modification of the linear energy-band structure of graphene for the multilayer case. Noise reduction in nano-scale few-layer graphene transistors is experimentally demonstrated and can be understood within this model as well.

Graphene, a single atomic layer of graphite, has attracted tremendous attention since it was first isolated by mechanical exfoliation in 2004[1]. Graphene exhibits an ultra-thin body, a unique energy band structure, superb electrical transport properties, and high mechanical and thermodynamic stability[2-6], which make it a promising candidate for future nanoelectronic devices. Besides single-layer graphene (SLG), two and few-layer graphene (FLG) are of interest for future device applications. While extensive studies have been carried out on the physical properties of SLG, less is known about the electrical properties of FLG structures. The focus of this article is on the charge and current distribution in FLG field-effect transistors (FETs) and the implications for the device performance.

It has been reported that graphene nanoribbon FETs fabricated on SiO$_2$ substrates with channel widths ≤ 30 nm exhibit high noise in the transfer characteristics at low temperatures and room-temperature[7-9]. The nanoribbon edge roughness, as a result of the top-down fabrication process is believed to be one of the contributing factors[10]. However, even for graphene FETs made from chemically derived and atomically smooth graphene nanoribbons, the current-voltage characteristics show an appreciable amount of noise at room-temperature[11]. What is common in all of the above devices is that a single-layer of graphene is used as the channel, a SiO$_2$ back-gate is employed to modulate the charge, and that the channel widths are 30 nm or less. The study presented here adds more insights on this topic. While we find that the noise in graphene nano-scale transistors is strongly affected by the number of graphene layers, the channel width, and the trapped charge in the SiO$_2$ substrate, consistent with the findings by Lin et al.[12], we do not have to assume any modification of the linear energy dispersion due to hybridization to explain our experimental results. The work presented here explores in particular the scaling impact of the graphene thickness on the device performance and indicates that the noise in aggressively scaled SLG FETs can be substantially reduced by employing FLG as the channel material. We present *the first* systematic experimental study on charge and current distribution in multilayer graphene (MLG) FETs. Based on our experimental findings, we propose a resistor network model describing the coupling between graphene layers including the impact of interlayer screening. The results of the model are consistent with previous work on intercalated graphite structures[13]. We propose that instead of SLG, two or few-layer graphene occurs to be the better choice for fabricating aggressively scaled nano-patterned graphene FETs for the benefits of lower noise and higher reliability.

Highly oriented pyrolitic graphite (HOPG) flakes were mechanically exfoliated on SiO$_2$/n+ Si substrates. A metal stack of Ti/Pd/Au (10 nm/30 nm/20 nm) was used to create the source/drain contacts employing e-beam lithography. A detailed description of a fabrication process for graphene nano-scale FETs can be found in reference (8). Figure 1(a) shows an SEM image of a nano-patterned graphene FET and the inset displays the device structure schematically. We chose to use $t_{ox}$ = 90 nm $SiO_2$ as the substrate since it provides a stronger contrast over the entire visible light wavelength range for thin graphene flakes to be clearly visible on $SiO_2$ under an optical microscope[14], and a more than three times stronger gate control of the graphene channel if compared with the commonly used 300 nm $SiO_2$ substrate. Figure 1(b) shows the room temperature transfer characteristics of three *broad area* graphene FETs, a SLG FET (1 ML), a bilayer graphene FET (2 ML), and a FLG FET (4 ML). All the curves are smooth and fairly symmetric about $V_{bg}$ = 0 V, which indicates the absence of unintentional doping and a clean/dry environment. The off-state is defined as the point with the minimum conductance ($V_{Dirac}$) and the on-state is defined where $V_{bg} - V_{Dirac}$ = -40 V. Apparently, the SLG FET provides the highest on-off ratio, the lowest off-conductance, the highest transconductance ($g_m = \partial I_d / \partial V_{gs}$, for $V_{ds}$ = const), and the sharpest transition at the Dirac point. However, the SLG FET becomes very noisy and displays varying device characteristics after being nano-patterned, while the nano-patterned FLG FET that was located on the same wafer, had the same geometry, went through the same processing steps, and was characterized under the same conditions shows much lower noise and better reproducibility, as shown in Figure 1(c). This experimental result is consistent with previous findings by other groups[7-9,12]. We thus have to conclude that the noisy behavior of the SLG FET is not due to the graphene edge roughness. In order to create the nanoribbons discussed above, we have employed a novel process that introduces a thin layer of PMMA (~50 nm thick) between the graphene and the HSQ e-beam resist. This approach enables a clean and non-destructive removal of HSQ after graphene channel definition using $O_2$ plasma RIE by dissolving the PMMA in acetone and lifting-off the HSQ rather than etching it. Due to the residual free removal of HSQ in this way, we believe that the apparent noise in the characteristics shown in Figure 1(c) is not caused by HSQ contamination.

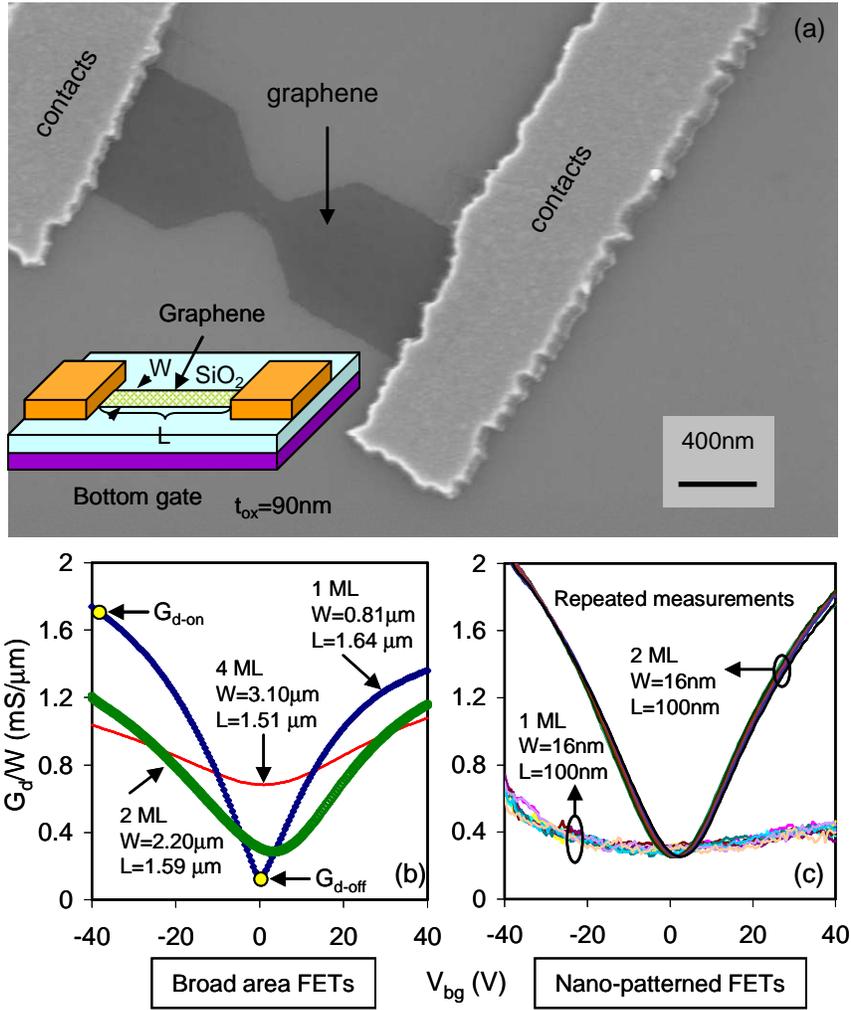

**Figure 1.** (a) SEM image of a nano-patterned graphene FET. The inset shows the device structure schematically. (b) Transfer characteristics of back-gated *broad area* graphene FETs with different graphene thickness. The yellow circles indicate on-state and off-state. Device dimensions are displayed in the plot. (c) Transfer characteristics of nano-patterned graphene FETs. Each set of curves show repeated measurements. The 1 ML graphene FET exhibits higher noise and poorer reproducibility than the 2 ML FET. All electrical measurements were performed at $T = 300$ K and $V_{ds} = 10$ mV.

To study the impact of the number of graphene layers in more details, we have fabricated ~ 40 broad area graphene FETs with thickness naturally varying between 0.35 nm and 3.7 nm, as determined by atomic force microscopy (AFM, Figure 2 (b)). We define 1 ML thickness as $d_{ML} \approx 0.35$ nm, which is the spacing between two adjacent graphene layers. Figure 2(a) shows the $I_{on}/I_{off}$ ratio of these devices vs. graphene thickness. The $I_{on}/I_{off}$ ratio indicates a 1/thickness dependence and reaches unity at around 10

ML (3.5 nm). The inset displays the same set of data in a log scale confirming $I_{on}/I_{off} \sim d^{-1}$. Figure 2(b) displays an AFM image of a graphene flake on $SiO_2$ substrate, along with its height profile to determine the thickness of the graphene flake. 1 ML and 2 ML graphene regions can be clearly distinguished.

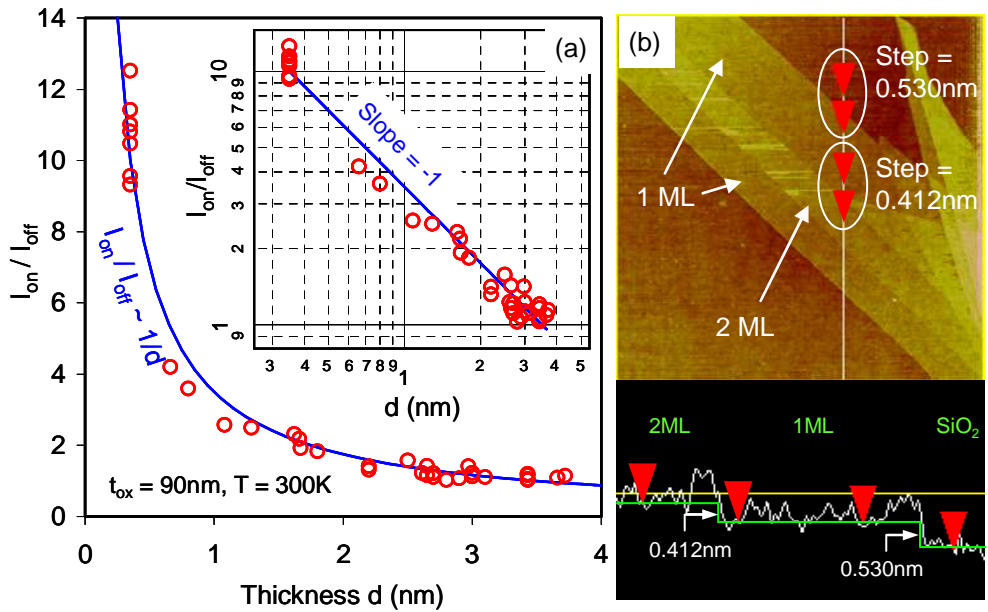

**Figure 2.** (a) $I_{on}/I_{off}$ ratio vs. graphene thickness of ~ 40 graphene FETs. The inset shows the same set of data in log scale, which clearly indicates a 1/thickness dependence of the $I_{on}/I_{off}$. (b) AFM image and height profile of a graphene flake on 90 nm $SiO_2$ substrate. 1 ML and 2 ML graphene regions as well as the step heights are indicated.

Figure 3 inspects in more details the individual thickness dependence of $G_{on}$ (triangles) and $G_{off}$ (circles). $G_{off}$ increases with thickness as expected assuming that an increasing number of graphene layers contribute to the total conductance. Surprisingly, $G_{on}$ shows a *decreasing* trend as more parallel graphene layers are added, which has never been reported before!

We employ a resistor network model shown in Figure 4(a) to explain our experimental observations. $G_1$, $G_2$…$G_N$ denotes the intralayer conductance of each graphene layer, and $R_{int}$ is the interlayer resistance due to coupling between the individual graphene sheets[15]. Coupling in our model captures the possibility of charge carriers tunneling from one graphene layer to the next. To assess this quantity experimentally, we have measured the resistivity of a "block" of HOPG (1cm x 1cm x 1.2 mm) along the c-axis. We determine a $\rho_c \approx 0.3$ Ω-m that translates into an interlayer resistance of $R_{int} = \rho_c \cdot d_{ML}/A \approx$ 105 Ω, assuming that the relevant area A for this process is given by the contact size of our devices. The intralayer resistance for a single graphene layer in the off-state ($R_{off}$) can be directly obtained from the transfer characteristic of a SLG FET, where $R_{off} = 5300$ Ω for the dimensions of our devices. Therefore, the ratio $R_{int}/R_{off}$ is estimated to be 0.02 ~ 0.2. This ratio is important for modeling of $I_{on}$ and $I_{off}$, as will become more obvious below.

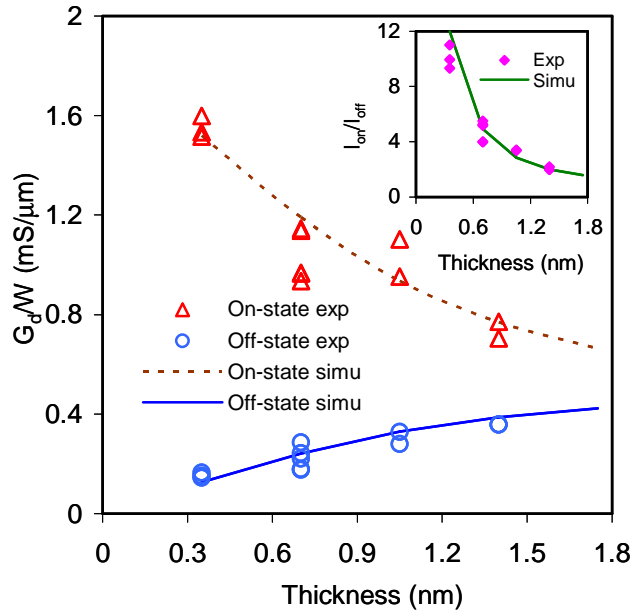

**Figure 3.** Normalized conductance vs. channel thickness for graphene FETs. "Dots" and "triangles" indicate experimental results, and dashed and solid lines indicate simulations. The simulation assumes $R_{int}/R_{off} = 0.05$ and $\lambda = 0.6$ nm. The inset shows the $I_{on}/I_{off}$ ratio vs. thickness of the same devices.

We make the following assumptions for the simulation of bottom-gated graphene FETs: (1) the source and drain contacts are connected to the top graphene layer only, since the contact metals were deposited at a relatively low temperature and did not undergo any high temperature annealing; (2) screening in the off-state is ignored since the screening length $\lambda$ exceeds the maximum considered graphene thickness[16]; (3) the maximum $I_{on}/I_{off}$ ratio is ~12, as experimentally observed in broad area SLG FETs (Figure 1(b), 1 ML).

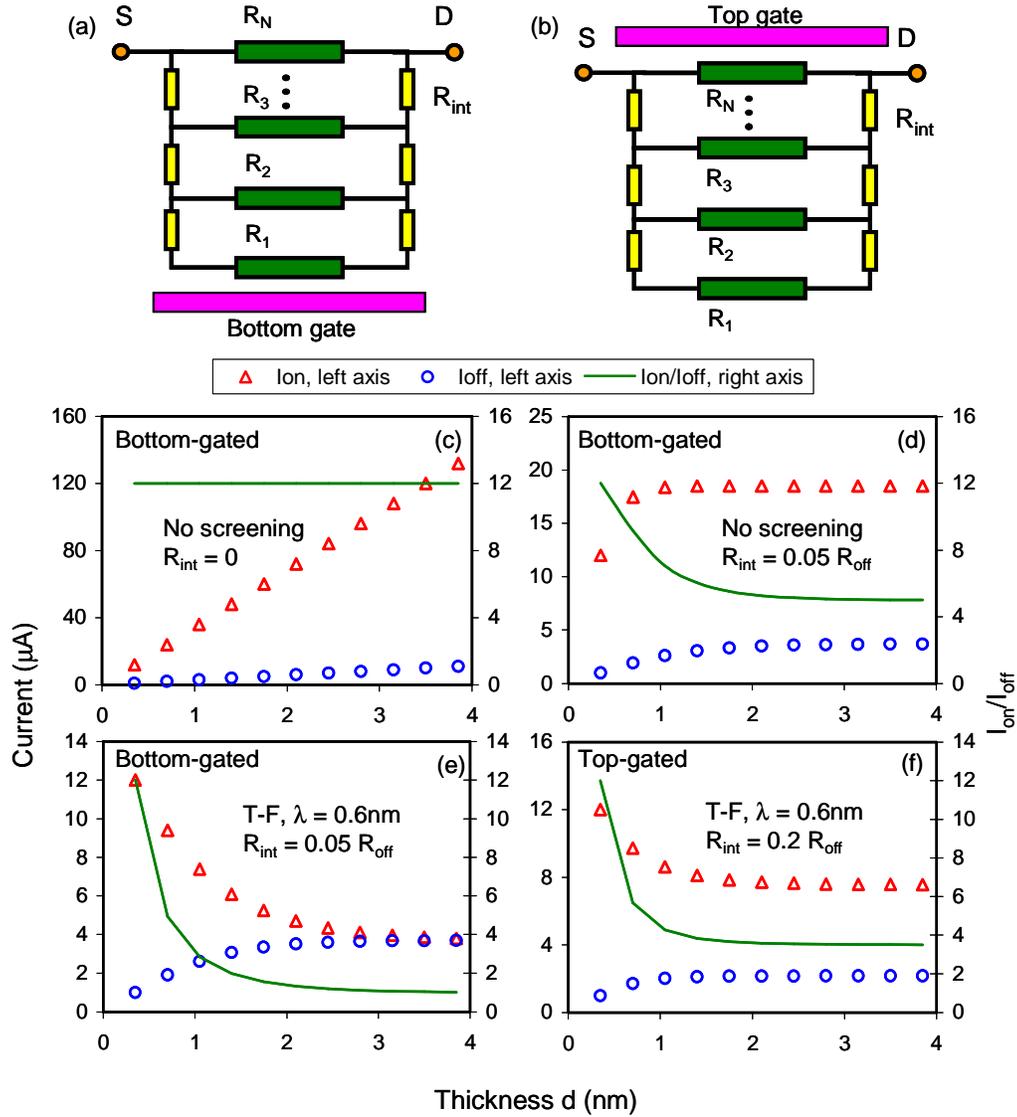

**Figure 4.** (a) A resistor network model representing a bottom-gated MLG FET. (b) A resistor network model representing a top-gated MLG FET. $R_i \equiv 1/G_i$, $i = 1, 2…N$. (c), (d), (e) Simulation results for bottom-gated MLG FETs using the network in (a), with assumptions indicated in each plot. (f) Simulation for a top-gated MLG FET using the network in (b).

Case 1, Figure 4(c): This is the simplest case, where we assume no screening in the on-state, i.e., constant $R_{on}$ for all layers, and $R_{int} = 0$. Both $I_{on}$ and $I_{off}$ increase linearly with thickness, and the $I_{on}/I_{off}$ ratio is constant, clearly not fitting our experimental data. Case 2, Figure 4(d): Here we ignore screening in the on-state and assume moderate interlayer coupling, $R_{int} = 0.05 \, R_{off}$. $I_{on}$ and $I_{off}$ both become constant after an initial increase with thickness. The initial increase of current is a result of the contribution from the additional parallel conduction paths (graphene layers), and the current saturation is due to $R_{int}$ preventing carriers to "dive" below a few layers from the top of the graphene stack. The corresponding $I_{on}/I_{off}$ ratio shows a decreasing trend with thickness, but the ratio is still ~5 at 10 ML, which is inconsistent with the experiment. If $R_{int}$ were to be made even larger, it will become increasingly difficult for the current to penetrate deeper graphene layers, and the $I_{on}$ and $I_{off}$ becomes nearly constant for all thicknesses. Again, this case does not capture our experimental observations.

Screening has to be considered to account for the rapid decrease of $I_{on}$ with thickness and the $I_{on}/I_{off}$ ratio. We have employed Thomas-Fermi (T-F) screening theory in our model to calculate the intralayer conductance in the device on-state. Earlier theoretical studies on screening in graphite[17] had found that a screened Coulomb potential of

$$\phi(z) \propto \frac{Q}{z} \exp\left(-\frac{z}{\lambda}\right) \tag{1}$$

(with Q being the charge under screening and z being the distance in the c-direction of the graphene stack) is an adequate description to capture the potential landscape in graphite. We further assume the following for the case studies that include screening: (1) each graphene layer has a minimum conductance regardless of the gating conditions[18]; (2) independent of the number of graphene layers (N) the total charge on the gate $Q_{gate}$ will be mirrored by the total induced charge in the channel region. The exact distribution of $Q_{gate}$ in the device on-state depends on the number of graphene layers and follows the potential variation according to equation (1). If $Q_1$ denotes the gate-induced charge in the bottom layer of the stack – the one closest to the gate, $Q_i$ the one in the i-th layer accordingly, and λ is the T-F screening length as a fitting parameter in our model, the gate-induced charge in each graphene layer can be written as

$$Q_i \propto \frac{1}{t_{ox} + r_i} \exp\left(-\frac{t_{ox} + r_i}{\lambda}\right) \tag{2}$$

$$\sum_{i=1}^{N} Q_i = Q_{gate} \tag{3}$$

(Here $r_i$ denotes the distance from the bottom of the graphene stack to the i-th layer.) Considering that the oxide thickness is much larger than the distance between graphene layers ($d_{ML}$) and that the 1/z dependence from equation (1) can be neglected if compared with the exponential dependence on z, the charge distribution can be calculated as

$$\frac{Q_i}{Q_{i-1}} \approx \exp\left(-\frac{r_i - r_{i-1}}{\lambda}\right) = \exp\left(-\frac{d_{ML}}{\lambda}\right) \equiv c \quad (4)$$

Therefore, the induced charge in the i-th layer can be derived from equation (3) and (4) as

$$Q_i = \frac{c^{i-1}}{\sum_{j=1}^{N} c^{j-1}} Q_{gate} \quad (5)$$

If we further assume that the on-state conductance induced by the gate in the i-th graphene layer is proportional to the induced charge in this particular layer, we find that the total on-state conductance of the i-th layer is

$$G_{on,i} = \frac{c^{i-1}}{\sum_{j=1}^{N} c^{j-1}} G_{gate} + G_{min} \quad (6)$$

where $G_{gate}$ is the conductance corresponding to the total gate charge, and $G_{min}$ is the finite minimum conductance for a graphene layer[18]. Note that transport in the device under investigation is neither ballistic, nor can we appropriately describe the $I_d$-$V_{gs}$ characteristics simply by considering one scattering mechanism - which is the motivation for utilizing $G_i \propto Q_i$ for simplicity. The off-state conductance in the i-th layer is

$$G_{off,i} = G_{min} \quad (7)$$

For SLG FETs on 90 nm $SiO_2$ substrate, the maximum $I_{on}/I_{off}$ ratio is around 12, and therefore, $G_{gate} \approx 11 \, G_{min}$.

Now we consider scenarios with the effects of both screening and interlayer coupling. Case 3, Figure 4(e): Assuming a T-F screening length of $\lambda = 0.6$ nm and $R_{int} = 0.05 \, R_{off}$, the on- and off-current can be calculated from equation (6) and (7) using the resistor network shown in Figure 4(a). As apparent from Figure 4(e), while $I_{on}$ decreases with thickness $I_{off}$ increases slowly. The $I_{on}/I_{off}$ ratio decreases rapidly with thickness and approaches unity around 10 ML. Obviously, both screening and interlayer coupling have to be considered to account for the experimental findings of the $I_{on}$, $I_{off}$, and $I_{on}/I_{off}$ dependence on thickness.

In order to determine $R_{int}$ and $\lambda$, we have fitted our on-conductance and the off-conductance simultaneously to the experimental data as shown in Figure 3. The simulation takes into account both the screening effect and the interlayer coupling effect. It turns out that $R_{int} = 0.05\ R_{off}$ and $\lambda = 0.6$ nm result in the best fit. The interlayer screening length obtained from the simulation agrees rather well with earlier theoretical studies that indicate a screening length of 0.7 nm for multilayer graphene[15], and 0.5 nm from transport measurements on graphite[16]. Moreover, the interlayer coupling parameter extracted from our simulation is within the predicted range of $R_{int}/R_{off} = 0.02 \sim 0.2$. It is the combined effects of screening and interlayer coupling that are responsible for the particular thickness dependence of the $I_{on}$, $I_{off}$, and on-off ratio for the bottom-gated MLG FETs, as experimentally observed.

For the case of SLG, there is neither screening nor interlayer coupling in the c-direction. Therefore, the entire SLG channel is very strongly coupled to the potential in the gate oxide. On one hand, this strong coupling is beneficial for obtaining a strong gate modulation, i.e., high $I_{on}/I_{off}$ ratio, high transconductance, and sharp transition at the Dirac point. On the other hand, any trapped charges in the oxide substrate or interface will lead to an appreciable potential variation in the SLG channel, which leads to a substantial amount of noise. The noise is usually negligible for broad area FETs, but for aggressively scaled nanostructures, the impact of the charges in the substrate can become severe, negating the advantages of SLG. Since the interlayer screening length is just over 1 ML thickness, the detrimental influence of potential fluctuations in the graphene due to oxide charge can be suppressed if 2 or 3 ML graphene is employed. Since noise in our picture is mainly a result of the current flowing through the bottom most graphene layer (next to the oxide), reducing its current contribution is an effective way to reduce the total noise in the device. In fact, the on-current through the bottom layer for a 3 ML graphene FET is only ~ 20% of the on-current through a 1 ML graphene FET of the same dimensions and at the same biasing conditions. Therefore, two or three-layer graphene may be a better choice for fabricating aggressively scaled back-gated nano-scale graphene FETs for the benefits of lower noise and higher reliability. In particular, we suggest that a somewhat reduced $I_{on}/I_{off}$ current ratio in few-layer graphene if compared to single-layer graphene structures can be compensated by the introduction of a bandgap through the formation of few-layer GNRs.

Last, we have employed our simulation tool to investigate top-gated MLG FETs, such as FETs fabricated from epitaxially grown graphene on SiC. The resistor network shown in Figure 4(b) is used for the simulation of top-gated MLG FETs. We assume the same screening length and weaker interlayer coupling for epitaxial graphene compared with exfoliated graphene due to an increased amount of material defects (such as rotational disorder) weakening the interlayer coupling in epitaxial graphene[19]. Figure 4(f) displays the $I_{on}$, $I_{off}$ and $I_{on}/I_{off}$ of the top-gated graphene FETs as a function of graphene thickness. We predict that the $I_{on}/I_{off}$ ratio for the top-gated structure only decreases mildly if compared with the bottom-gated case when the graphene layer thickness is increased. This is the case since the top a few graphene layers are the ones that "see" most of the gate modulation and are at the same time the ones closely connected with the source and drain contacts. Furthermore our simulation suggests that top-gated devices should be built on few layer graphene – rather than single layered graphene – in this way allowing on one hand for a sufficient gate modulation between the on- and off-state while providing sufficient screening of charge related substrate effects. It is also encouraging to see both the $I_{on}$ and $I_{off}$ are nearly invariant with graphene thickness if the epitaxial graphene is more than 3 ML thick, which relieves the rigorous requirement for growing uniform epitaxial graphene over a large area with atomic smoothness. We suggest that material growth studies should focus on the control of interlayer coupling strength to improve the $I_{on}/I_{off}$ ratio in a robust fashion. Initial experimental evidence indicates that top-gated epitaxially grown MLG FETs can provide a decent $I_{on}/I_{off}$ ratio with rather thick (~ 2nm) graphene channel[20], which corroborates with our predictions.

In conclusion, noticing the importance of noise impacting the performance of nano-patterned graphene FETs, we have performed *the first* systematic experimental study on multilayer graphene FETs to address the charge and current distribution in MLG FETs. We have developed a model describing the coupling between graphene layers including interlayer screening effects. The model *does not* involve modification of the linear energy band structure for multilayer graphene. Simulation results using our model are in good agreement with the experimental findings as well as previous studies on intercalated graphite structures[13,16,18]. Noise reduction in nano-scale FLG transistors is experimentally demonstrated and can be understood within this model. Moreover, we predict that the demand for extremely thin and uniform epitaxial graphene (for decent $I_{on}/I_{off}$ ratio) may not be entirely justified for transistor application. In particular, we propose two or three-layer graphene to be the ideal candidate for aggressively scaled nano-patterned graphene transistors.

ACKNOWLEDGMENT. The authors thank Prof. Mark Lundstrom, Prof. Supriyo Datta and Christian Sandow for insightful discussions. This work was supported by NRI and Intel.